# Domain Topology and Domain Switching Kinetics in a Hybrid Improper Ferroelectric

F.-T. Huang[1], F. Xue[2], B. Gao[1], L. H. Wang[3], X. Luo[3], W. Cai[1†], X. Lu[4], J. M. Rondinelli[4], L. Q. Chen[2], and S.-W. Cheong[1,3*]

[1] Rutgers Center for Emergent Materials and Department of Physics and Astronomy, Rutgers University, Piscataway, New Jersey 08854, USA

[2] Department of Materials Science and Engineering, The Pennsylvania State University, University Park, Pennsylvania 16802, USA

[3] Laboratory for Pohang Emergent Materials and Max Plank POSTECH Center for Complex Phase Materials, Pohang University of Science and Technology, Pohang 790-784, Korea

[4] Department of Materials Science and Engineering, Northwestern University, USA

*email: sangc@physics.rutgers.edu

†Present address: Chongqing University of Science and Technology, Chongqing, 401331, China

**Charged polar interfaces such as charged ferroelectric domain walls or heterostructured interfaces of ZnO/(Zn,Mg)O and LaAlO$_3$/SrTiO$_3$, across which the normal component of electric polarization changes suddenly, can host large two-dimensional conduction. Charged ferroelectric walls, which are energetically unfavorable in general, were found to be mysteriously abundant in hybrid improper ferroelectric (Ca,Sr)$_3$Ti$_2$O$_7$ crystals. From the exploration of antiphase boundaries in bilayer-perovskites, we discover that each of four polarization-direction states is degenerated with two antiphase domains, and these eight structural variants form a Z$_4\times$Z$_2$ domain structure with Z$_3$ vortices and five distinct types of domain walls, whose topology is directly relevant to the presence of abundant charged walls. We also discover a zipper-like nature of antiphase boundaries; they are the reversible creation/annihilation centers of pairs of two types of ferroelectric walls (and also Z$_3$-vortex pairs) in 90° and 180° polarization switching. Our results demonstrate the unexpectedly rich nature of hybrid improper ferroelectricity.**

Over the last decade, two-dimensional (2D) conduction in heterostructured interfaces with polar discontinuity[1,2] or compositionally homogeneous charged interfaces such as charged ferroelectric domain walls (FE DWs)[3-10] has attracted enormous attention for emergent phenomena and new materials functionalities. However, un-desirable chemical/structural complexity such as ionic diffusion, oxygen vacancies, or structural/strain variations near the interface results in equivocal interpretation of the origin of 2D conduction[11-13]. In parallel, significant efforts have been launched to investigate the 2D conduction at charged FE DWs, which are well defined in the atomic scale. The conducting FE DWs were observed in chemically homogeneous thin films, e.g., $Pb[Zr_xTi_{1-x}]O_3$[14], $BiFeO_3$[10] and $Bi_{0.9}La_{0.1}FeO_3/SrRuO_3$ heterostructure[9], and bulk single crystals, e.g., $LiNbO_3$[4], $BaTiO_3$[5], and $Er(Ho)MnO_3$[6,7]. These conducting FE DWs are intrinsically unstable due to large energy cost[8,15], often pinned by chemical defects[4], and sporadic[5] and artificially created with external voltage[9]. However, charged DWs, some of which are highly conducting, were found to be abundant in the hybrid improper ferroelectric $(Ca,Sr)_3Ti_2O_7$[3].

Ordering phase transition in condensed matter can be accompanied by directional variants and antiphase boundaries (APBs). Directional-variants result in domains with different directional order parameters. The role of APBs[16,17] on materials functionalities has been well recognized[18-22]. In particular, the discovery of strong interaction or interlocking nature of APBs with ferroic orders in diverse functional materials opens new grounds for materials research. Examples include the ferromagnetic coupling in Heusler alloys[23], the reduced spin polarization in half-metal magnetites[24], the duality nature of DWs and topological defects in hexagonal manganites[25], ferroelectric APBs in a nonpolar matrix[16], and conducting and ferromagnetic walls of antiferromagnetic domains in pyrochlore iridates[26].

Hybrid improper ferroelectricity (HIF), a phenomenon involving polarization induced by a hybridization of two non-polar lattice instabilities, offers great promise toward the realization of room-temperature multiferroelectricity[27-31]. The key idea is to design new materials in which ferroelectricity and (anti)ferromagnetism can be coupled by the same lattice instability, therefore providing an indirect but strong coupling between polarization and magnetism[27-29,31,32]. Examples of compounds with HIF include the double-layered Ruddlesden-Popper (RP) perovskites with the chemical formula of $A_3B_2O_7$ (Fig. 1, $A^{2+}$ = alkali metal; $B^{4+}$ = transition metal)[3,27,32]. Unexpectedly, charged FE DWs, some of which are highly conducting, were also found to be mysteriously abundant in the recently discovered RP-type HIF $(Ca,Sr)_3Ti_2O_7$ crystals[3]. In order to unveil the origin of these abundant charged FE DWs, we have explored the complete connectivity of DWs in $Ca_{2.55}Sr_{0.45}Ti_2O_7$ (CSTO; FE Tc≈790 K) and $Ca_3Mn_{1.9}Ti_{0.1}O_7$ (CMTO; FE Tc≈360 K) single crystals with in-plane polarization along the pseudo-tetragonal [110] directions[3,30,33], particularly with mapping of APBs using Transmission Electron Microscopy (TEM). Note that APBs are invisible in Piezoresponse Force Microscopy (PFM)[34], which is usually a good method to map out FE domain configurations. Phase-field simulations[35] were also conducted to understand the origin of domain configurations in CMTO and CSTO. Our results reveal that the formation of a unique $Z_4xZ_2$ domain topology with $Z_3$ vortices is responsible for the presence of abundant charged FE DWs in CSTO. In addition, we have also investigated the kinetics associated with polarization switching in CSTO, understanding of which is crucial for developing precise control of conducting FE walls.

**Results**

**$Z_4 \times Z_2$ domain structure with eight different states.** Figure 1a shows two characteristic lattice modes in $A_3B_2O_7$: first, the $BO_6$ octahedra in-phase rotations in either clockwise (the sign of the rotation is +) or counterclockwise (-) about a $[001]_{tetra}$ direction (denoted as $a^0a^0c^+$ in the Glazer notation[36] or the $X_2^+$ mode), and second, the $BO_6$ octahedral tilting about two $<110>_{tetra}$ axes, i.e. apical oxygen-motions toward the $1^{st}$ - $4^{th}$ quadrants (denoted as $a^-a^-c^0$ or the $X_3^-$ mode). Below the phase transition, the $X_3^-$ mode adopts one of the four tilts (1-4) accompanying the $X_2^+$ mode with + or - rotations into a combined distortion pattern of $a^-a^-c^+$ having eight degenerate states, which we label as 1±, 2±, 3±, and 4± (the complete structures are shown in Fig. 1b-c). Figure 1b shows the 1+ state projected along the $[001]_{tetra}$ and $[010]_{tetra}$ directions. For example, the apical oxygen of the blue circled octahedron moves toward the $1^{st}$ quadrant (purple arrow) with a clockwise rotation (black curved arrow) to form a 1+ state. A trilinear coupling among the $X_2^+$ mode, $X_3^-$ mode and the polar $X_5^-$ mode (A-site displacement) yields four FE polarizations parallel or antiparallel to the two $<110>_{tetra}$ tilting axes[3,27]. Emphasize that each polarization direction is associated with two degenerate states, e.g. the 1+ and 3- polarizations point in the same direction (red and light-red arrows in Fig. 1b-c), a consequence of both nonpolar order parameters ($X_3^-$ and $X_2^+$) changing signs. These four polarization directional variants with two-fold degeneracy form the basis of the $Z_4xZ_2$ domain structures in the HIF $A_3B_2O_7$. Figure 2a shows a structural model of a APB (green line) between the 1+ and 3- states, at which the orthorhombic unit cells labelled by green dotted lines demonstrate the discontinuation of octahedral tilting (purple arrows) and rotation (+ and -) at the wall. APBs might exist in $A_3B_2O_7$, but behave hidden in the PFM images since the two domains give the same piezo-response[3].

**Z₃ vortices in the Z₄×Z₂ domain structure of (Ca,Sr)₃Ti₂O₇.** Polarized optical microscope (POM) images on CSTO and CMTO crystals both clearly exhibit the existence of orthorhombic twins, i.e. orthorhombically-distorted ferroelastic (FA) domains. In addition, in-plane PFM studies show the intriguing FE domains comprising abundant meandering head-to-head and tail-to-tail charged DWs[3] (Supplementary Figs. S1). Compared with PFM, dark-field TEM (DF-TEM) under systematic controlled diffraction conditions allows us to light up domains induced directly by local structural deformations[37]. Figures 2b-d show a series of DF-TEM images taken along the $[001]_{tetra}$ direction using superlattice peaks $g_1^{\pm} = \pm 3/2(1, 1, 0)_{tetra}$ parallel to the polar axis within a single FA domain. Three domains (i-iii) in Figs. 2b-d, in which domains i and ii reveal the same domain contrast but opposite to domain iii in contrast, demonstrate the existence of antiphase domains i and ii, and an APB between them. Three FE domains including two antiphase domains merging at one vertex point are well illustrated in Figs. 2b-2d. Since a sense of rotation along the merging three domains is defined in the phase space (see below), the vertex structure can be called as a Z₃ vortex. Note that the relative polar $\pm a_{orth}$ directions can be identified from the related electron diffraction patterns but the absolute polarization direction cannot be. Thus, once the polarization direction is chosen for one domain, then the polarization directions in other domains can be fully assigned without ambiguity. Evidently, the existence of APBs remains unchanged even if the assignment of polarization direction is reversed. Figure 2e depicts one possible assignment with 1+ and 3- antiphase domains. The APB between these 1+ and 3- domains accompanies the sign change of both rotation and tilting (Fig. 2a). The presence of an APB and a Z₃ vortex also suggests the existence of pure rotation $(a^0a^0c^+)$-driven DWs and pure tilting $(a^-a^-c^0)$-driven DWs. We define a tilting-type $FE_t$ DW (rotation-type $FE_r$ DW) as the wall between adjacent FE domains having opposite $a^-a^-c^0$ tilting $(a^0a^0c^+$ rotation) but identical $a^0a^0c^+$ rotation $(a^-a^-c^0$ tilting). The structural details of $FE_t$ and $FE_r$ DWs are shown in Supplementary Figs. S2-S3. Note that the Z₃ vortex in Fig. 2e consists of three distinct walls: $FE_r$ (red-dotted), $FE_t$ DWs (red-solid) and APB (green line).

Figure 3a shows a mosaic of DF-TEM images covering three ferroelastic (i.e. orthorhombic twin; FA(i) and FA(ii)) regions in a CSTO crystal. In the DF-TEM image obtained using orthorhombic superlattice peaks $g_1^{\pm}$ contributed from the FA(i) domain marked by red and blue circles in Fig. 3b, the neighboring FA(ii) regions exhibit a dark contrast (Fig. 3a). Aside from the major contrast between FA(i) and FA(ii), a profound self-organized Z₃-vortex network is clearly visible within the FA(i) region. The inset of Fig. 3a shows a schematic overlying the middle rectangular area, in which a pair of $Z_3$ vortices is linked by an APB (green line). Boundaries between the 3- (pink) and 3+ (blue) domains form broad contrast walls, identified as $FE_r$ DWs (red-dotted lines in the inset of Fig. 3a). By comparing the domain contrasts (Fig. 3c), wall features and the neighboring FA domains (Supplementary Figs. S4) obtained from our DF-TEM images, we marked out the full assignment of domain states and wall-types of Fig. 3a in the Supplementary Figs. S4.

**Domain topology of $Ca_3(Mn,Ti)_2O_7$.** We also grew high-quality CMTO single crystals and confirmed the presence of polar domains in the same polar space group ($A2_1am$) as $Ca_3Ti_2O_7$ at 300 K. Figure 3f shows a POM image of surprisingly irregular FA DWs on the cleaved (001) surface, distinct from the prototypical straight FA DWs (i.e. orthorhombic twin walls) in CSTO. Our DF-TEM images (Fig. 3d-e) demonstrate consistently the presence of irregular twin patterns. The FA(i) domain is excited when the red-circled $g_1^+$ spot was used for imaging (Fig. 3d), and it turns to deep-dark contrast when the orthogonal yellow-circled spot was excited (Fig. 3e). Inside each FA domain, there exist 180º-type FE domains; for example, bright and gray contrast domains in Figs. 3d-e and Supplementary Figs. S5. The domain configuration in the inset of Fig. 3d displays the presence of $Z_3$ vortices with three domains merging at the vortex cores, which exist within a FA domain, as well as at boundaries between FA domains. Thus, the configuration of $Z_3$-vortex domains seems universal in the HIF $A_3B_2O_7$, despite the existence of eight possible structural variants. Note that there exist two types of FA DWs: ferroelastic tilting DWs ($FA_t$ DW) between states in the same rotation e.g. the 1+ & 2+ or 3- & 2- (blue-dotted lines in Fig. 3d and Supplementary Figs. S2), and ferroelastic tilting+rotation ($FA_{tr}$) DWs between, e.g., the 2+ & 3- states (Supplementary Figs. S2). A single tilting of either $a^-a^0c^0$ or $a^0a^-c^0$ type may occur at $FA_t$ and $FA_{tr}$ DWs (white arrows in Supplementary Figs. S2). $FA_{tr}$ DWs seem naturally accompanied with a complete octahedral-rotation frustration at the walls, implying a higher energy of $FA_{tr}$ DWs than that of $FA_t$ DWs (Supplementary Figs. S3).

**Phase-field simulations.** We also employed the phase-field method[35] to investigate HIF $A_3B_2O_7$, as shown in Figs. 4a-4b. The details of the simulations are given in Methods. The eight states are represented by different colors. $Ca_3Ti_2O_7$ exhibits straight FA DWs (Fig. 4a), consistent with the experimental observation[3] (Supplementary Figs. S1). Across the FA DWs, a dark color tends to become another dark color, e.g. a red (1+) color tends to change to a green (2+) color, rather than to a light green (4-) color (Fig. 4a, circle b). Similarly, two light colors tend to be close with each other across the FA DWs (Fig. 4a, circle c). These indicate that the rotation order parameter tends to be unchanged across the FA DWs[38], which is consistent with the previous discussion on the low energy of $FA_t$, compared with $FA_{tr}$ DWs. Supplementary Figs. S3 show the oxygen positions of the eight states relative to the tetragonal position. With the assumption that a wall going through a tetragonal-like state with zero polarization costs more energy, the energy hierarchy among the five DWs can be estimated to be $FA_t \leq FE_r \leq FE_t < FA_{tr} \leq APB$. The statistics of DW lengths obtained from phase-filed simulation results give a ratio of $FE_r:FA_t:FE_t:APB:FA_{tr}=34:28:28:8:2$ in $Ca_3Ti_2O_7$ (Fig. 4a) and $FA_t:FE_r:FA_{tr}:FE_t:APB=52:22:12:7:7$ in $Ca_3Mn_2O_7$ (Fig. 4b). A comparably large population of $FE_t$, $FE_r$ and $FA_t$ walls, especially in $Ca_3Ti_2O_7$, suggests that APB and $FA_{tr}$ walls belong to the higher energy set than others. Experimentally, within a limited number of domain walls that we have observed, CSTO exhibits an 82% population of the lower energy set ($FE_r$, $FE_t$ and $FA_t$ walls), while CMTO shows a 64% population. The low population in CMTO is likely related with the existence of irregular FA domains in CMTO. Emphasize that even though an energy hierarchy may exist, we do observe experimentally and theoretically all five kinds of DWs. The energy diagram for the HIF $A_3B_2O_7$ such as Fig. 4c can be constructed with all eight states (vertices) and all five kinds of walls (edges connecting two vertices). Note that in Fig. 4c, only a small part of edges are shown, and each vertex is connected to all of the rest vertices through edges in the full energy diagram. The $a$, $b$ and $c$ loops in Fig. 4c correspond to $Z_3$ vortices in the Fig. 4a-4b, respectively. The presence of only $Z_3$ vortices indicates that the energy difference among five types of DWs is not large, so the lowest-energy vortex defect is always $Z_3$-type. Note that if, e.g. APB is associated with much higher energy than others, then APB will be fully avoided, and $Z_4$-type vortex defects such as 1+/3+/3-/1-/1+ can occur, but we observe only $Z_3$-type vortices. In low-magnification TEM images, we sometimes observe vortices looking like $Z_4$-type. However, all these likely $Z_4$-type vortices turn out to be pairs of closely linked $Z_3$ vortices with inclined broad DWs, as shown in Fig. 4d-f. The energy diagram (Fig. 4c) is, in fact, a hyper-tetrahedron in 7 dimensions, which has eight vertices and only triangular faces. Each triangular face in this hyper-tetrahedron corresponds to a $Z_3$ vortex. All possible configurations of $Z_3$-votex are shown in Supplementary Figs. S6. The configuration of $Z_3$ vortex domains seems universally adopted in HIF $A_3B_2O_7$ compounds. Note that FA DWs in $Ca_3Ti_2O_7$, nucleated from a high-temperature tetragonal phase ($I4/mmm$)[33,39], tend to be straight, while FA DWs in Ca₃Mn₂O₇, nucleated from the *Acaa* phase[30] below ~360 K, are irregular. (Supplementary Figs. S5). A phase-field simulation for the nucleation and growth of the ferroelectric $A2_1am$ domains from the *Acaa* matrix is shown in Supplementary Movie I.

**Domain switching kinetics.** In-situ poling results on CSTO using a DF-TEM technique unveil intriguing domain switching kinetics, which can be understood in terms of the creation and annihilation of Z₃ vortex-antivortex (V-VA) pairs. In-situ poling is achieved utilizing fast positive charging[40,41] induced by focusing the electron beam (~300 nm in diameter) of TEM at a thin and local area, and the slow reduction of effective electric fields, due to charge dissipation, occurs after removing the focused beam. Thus, in order to observe the in-situ poling process, DF-TEM images are obtained immediately after defocusing the electron beam, before the charges are completely dissipated (on the order of 30 minutes). Figures 5a-d show a Z₃ V-VA pair (blue and cyan circles) within one FA domain behaving coherently with in-situ poling. In-situ poling at a 5 o'clock position near the crystal edge induces a direct 180° polarization reversal of a 4- (light green) domain to a 4+ (yellow) domain (Fig. 5b), which is accompanied by the creation of a V-AV pair. The induced 4+ domain shrinks slowly after defocusing the electron beam (from Fig. 5b to 5d). Eventually, the induced 4+ domain disappears, and at the same time, the V-AV pair annihilates. This result demonstrates a 180° polarization reversal associated with the creation or annihilation of a Z₃ V-AV pair. Furthermore, the results in Figs. 5a-d reveal that APBs act as nucleation reservoirs for the Z₃ V-AV pair creation and annihilation (Supplementary Figs. S7 and Movie II for more details). Figure 5e illustrates that in a Z₃ V-AV pair creation process, a segment of a APB becomes two FE DWs (one FE$_r$ DW and the other FE$_t$ DW); FE$_t$ DWs with a larger energy than that of FE$_r$ tend to be pinned at the original APB location while FE$_r$ DWs tend to be mobile. This observation is in accordance with the energy hierarchy of FE$_r$ ≤ FE$_t$ < APB discussed earlier. We also studied electron beam-induced poling in different directions. The switching process depends significantly on the electric field orientation (Supplementary Figs. S7). For example, Fig. 5g shows another region with two antiphase 4+ (yellow) and 2- (light yellow) domains with slightly different bright contrasts located next to an APB (green line). Interestingly, when the electron beam is focused on the specimen edge away from FA DWs (blue lines) and an electric field (red arrow in Fig. 5h) perpendicular to the original polar axis is induced, a 90° polarization switching from a bright 2- (light yellow) to dark 3- (pink) triangular domain is observed. The induced 3- domain returns slowly to the initial 2- state with charge dissipation. This process is involved with the splitting or coalescence of an APB to two ferroelastic DWs; one $FA_{tr}$ DW and the other $FA_t$ DW (Fig. 5f). The created $FA_{tr}$ DW with high energy stays at the original APB location while the created $FA_t$ DW with low energy tends to be mobile.

**Discussion**

We do not observe any evidence for going through an intermediate 90° (180°) polarization state in the process of 180° (90°) polarization switching. We emphasize that the coherent domain wall network of $Z_4xZ_2$ domains with $Z_3$ vortices and some highly-curved walls (Supplementary Figs. S7-S8) leads to the presence of charged domain walls and APBs. The presence of $Z_3$-vortices instead of the formation of antiparallel domains separated by neutral walls also indicates that a moderate energy difference among five types of DWs. In particular, those APBs serve as primary nucleation centers for 180° and 90° polarization switching, and the presence of mobile $n$-type charge carriers, screening the polar discontinuity at charged domain walls, are responsible for the large conduction of head-to-head domain walls[3]. Note that the APBs with the discontinuity of both octahedral tilting (t) and rotation (r) costs more energy and one ABB splits into two ferroelectric/ferroelastic walls under an external electric field, so an APB becomes a nucleation center of a new poled domain (Figure 5). This can also happen in high-energy $FA_{tr}$ DWs as shown in the Supplementary Figs. S8. This zipper-like splitting of high-energy DWs accompanies the emergence of $Z_3$ V-AV pairs. The high-energy APB and $FA_{tr}$ DWs dominate the nucleation controlled kinetics of polarization flipping, while the low-energy $FE_r$ and $FA_t$ DWs tend to move steadily with an external electric field, which can be responsible for the domain walls motion kinetics. Emphasize that unlike expected minor roles of APBs with just translation phase shift, APBs dominate the ferroelectric polarization switching in hybrid improper ferroelectric $(Ca,Sr)_3Ti_2O_7$. These unexpected discoveries of the role of APBs and the domain topology relevant to the presentence of abundant conducting domain walls should be further investigated for deeper understanding and nano-engineering of the domains and domain walls in hybrid improper ferroelectrics.

**Methods**

**Sample preparation.** Single-crystalline $Ca_{2.55}Sr_{0.45}Ti_2O_7$ (CSTO) and $Ca_3Mn_{1.9}Ti_{0.1}O_7$ (CMTO) were grown by using optical floating zone methods. For polycrystalline $Ca_{3-x}Sr_xTi_2O_7$ ($Ca_3Mn_{2-}$ $_x$Ti$_x$O$_7$) feed rods, stoichiometric CaCO$_3$, SrCO$_3$ and TiO$_2$ (MnO$_2$) were mixed, ground, pelletized and sintered at 1350-1550 °C for 30 h. Substituting Sr into the Ca site in Ca$_3$Ti$_2$O$_7$ induces the reduced size of FA domains suitable for transmission electron microscopy (TEM) studies. It was very difficult to grow high-quality single crystals of pure Ca$_3$Mn$_2$O$_7$, but the slight substitution of Ti into the Mn site of Ca$_3$Mn$_2$O$_7$ stabilizes crystal growth without changing the relevant physics of Ca$_3$Mn$_2$O$_7$. Crystals are highly cleavable, and were cleaved in air for optical microscopy observation. Transparent amber colored CSTO single crystals and non-transparent dark-blue colored CMTO present a similar polar orthorhombic symmetry (SG #36, $A2_1am$)[3,30,33]. Crystal structure and lattice parameters were examined by x-ray diffraction with a Philips XPert powder diffractometer and the general structure analysis system (GSAS) program. CMTO possesses a twice larger ferroelastic distortion (i.e., orthorhombicity defined by (a-b)/(a+b)*100%, ~0.05 %) than Ca$_3$Ti$_2$O$_7$ (~0.02 %). Cycling the CMTO sample temperature through Tc leads to a completely different irregular FA pattern, indicating that the domain formation is not simply due to pinning by disorder such as chemical defects or dislocations.

**Dark-field TEM measurement.** Specimens for dark-field transmission electron microscopy (DF-TEM) studies were fabricated on Ca$_3$Ti$_2$O$_7$ (CTO), Ca$_{2.55}$Sr$_{0.45}$Ti$_2$O$_7$ (CSTO) and Ca$_3$Mn$_{1.9}$Ti$_{0.1}$O$_7$ (CMTO) single crystals (~1 x 2 x 0.1 mm$^3$ in size) by mechanical polishing, followed by Ar-ion milling and studied using a JEOL-2010F TEM. We observed vortex domains by DF-TEM imaging taking two diffraction vectors: (1) superlattice $g_1^\pm = \pm 3/2(1, 1, 0)_{tetra} = \pm (3, 0, 0)_{orth}$ spots, parallel to the polar axis in $[001]_{tetra}$ zone and (2) superlattice $g_2^\pm = \pm 3/2(-1, 1, 2)_{tetra} = \pm (0, -3, 3)_{orth}$ spots, perpendicular to the polar axis in $[1, -1, 1]_{tetra}$ zone, 15° tilting from $c$-axis. Note that one TEM specimen can include up-to-a-hundred FA domains for observations, and we have examined two CSTO, one CTO and two CMTO TEM specimens. Although the width of FA domains varies, our conclusion on the $Z_4$x$Z_2$ domain structure with $Z_3$-vortex patterns and five types of domain walls is universal in all specimens that we have observed. We also found that the FA domain size depends little on various heat treatment conditions in both CSTO and CMTO.

**Phase-field modeling.** To describe the distortion relative to the high symmetry phase with space group *I4/mmm*, three sets of order parameters are used, i.e. $\varphi_i (i = 3)$ for the oxygen octahedral rotation around the $x_3$ axis, and $\theta_i (i = 1,2)$ and $P_i (i = 1,2)$ for the octahedral tilt and polarization component along the $x_i (i=1,2)$ pseudocubic axis, respectively[23,42]. The total free energy density can be expressed by

$$f = \alpha_{ij}\varphi_i\varphi_j + \alpha_{ijkl}\varphi_i\varphi_j\varphi_k\varphi_l + \beta_{ij}\theta_i\theta_j + \beta_{ijkl}\theta_i\theta_j\theta_k\theta_l + t_{ijkl}\varphi_i\varphi_j\theta_k\theta_l + d\varphi_3(\theta_1 P_1 + \theta_2 P_2) + \gamma_{ij}P_iP_j$$
$$+ \kappa_{ijkl}\frac{\partial\varphi_i}{\partial x_j}\frac{\partial\varphi_k}{\partial x_l} + \delta_{ijkl}\frac{\partial\theta_i}{\partial x_j}\frac{\partial\theta_k}{\partial x_l} + g_{ijkl}\frac{\partial P_i}{\partial x_j}\frac{\partial P_k}{\partial x_l} + \frac{1}{2}c_{ijkl}(\varepsilon_{ij}-\varepsilon_{ij}^0)(\varepsilon_{kl}-\varepsilon_{kl}^0) - \frac{1}{2}E_i P_i, \tag{1}$$

where $\alpha_{ij}$, $\alpha_{ijkl}$, $\beta_{ij}$, $\beta_{ijkl}$, $t_{ijkl}$, $d$, and $\gamma_{ij}$ are the coefficients of Landau polynomial, $\kappa_{ijkl}$, $\delta_{ijkl}$, and $g_{ijkl}$ are gradient energy coefficients, $c_{ijkl}$ is the elastic stiffness tensor, $\varepsilon_{ij}$ and $\varepsilon_{kl}^0$ are the total strain and eigen strain, and $E_i$ is the electric field given by $E_i = -\varphi_{,i}$ with $\varphi$ the electrostatic potential. Note that $\gamma_{ij} > 0$, and the term $d\varphi_3(\theta_1 P_1 + \theta_2 P_2)$ determines that $(P_1, P_2, 0)$ and $(\theta_1, \theta_2, 0)$ are parallel or antiparallel, dependent on the sign of $\varphi_3$. The eigen strain is related to the order parameters through $\varepsilon_{ij}^0 = \lambda_{ijkl}\varphi_k\varphi_l + h_{ijkl}\theta_k\theta_l$, where $\lambda_{ijkl}$ and $h_{ijkl}$ are coupling coefficients. Here the coupling between polarization and strain is ignored, since the secondary order parameter polarization is always parallel or antiparallel to the octahedral tilt order parameter. The Landau polynomial is expanded based on group theory analysis[42], and the related coefficients are obtained by first-principles calculations[38,43-48] (Supplementary Table 1-3). Anisotropic properties are assumed in gradient energy coefficients $\kappa_{ijkl}$ and $\delta_{ijkl}$, i.e. $\kappa_{iiii} << \kappa_{ijij}$, $\delta_{iiii} << \delta_{ijij}$[38]. The phase-field equations are solved with the initial condition of zero plus a small random noise for the order parameter components[49]. Periodic boundary conditions are employed along the three directions. The system size is 1024 $\Delta x \times 1024$ $\Delta x \times 1 \Delta x$, and the grid spacing is $\Delta x = 0.30\,nm$.

**Ackowledgments**

The work at Rutgers was supported by the Gordon and Betty Moore Foundation's EPiQS Initiative through Grant GBMF4413 to the Rutgers Center for Emergent Materials. The work at Penn State is supported by the NSF MRSEC under Grant No. DMR-1420620 (FX).

**Author contributions**

F.-T. H. conducted the TEM experiments and wrote the paper. F. X., X. L., J. M. R. and L. Q. C. performed the modelling, G. B., L. H. W. and X. L. synthesized single crystals, and W. C. performed the IP-PFM. S-W.C. initiated and supervised the research.

**Additional information**

The authors declare no competing financial interests. Supplementary information accompanies this paper on xxx. Reprints and permissions information is available online at xxx. Correspondence and requests for materials should be addressed to S.-W. C.

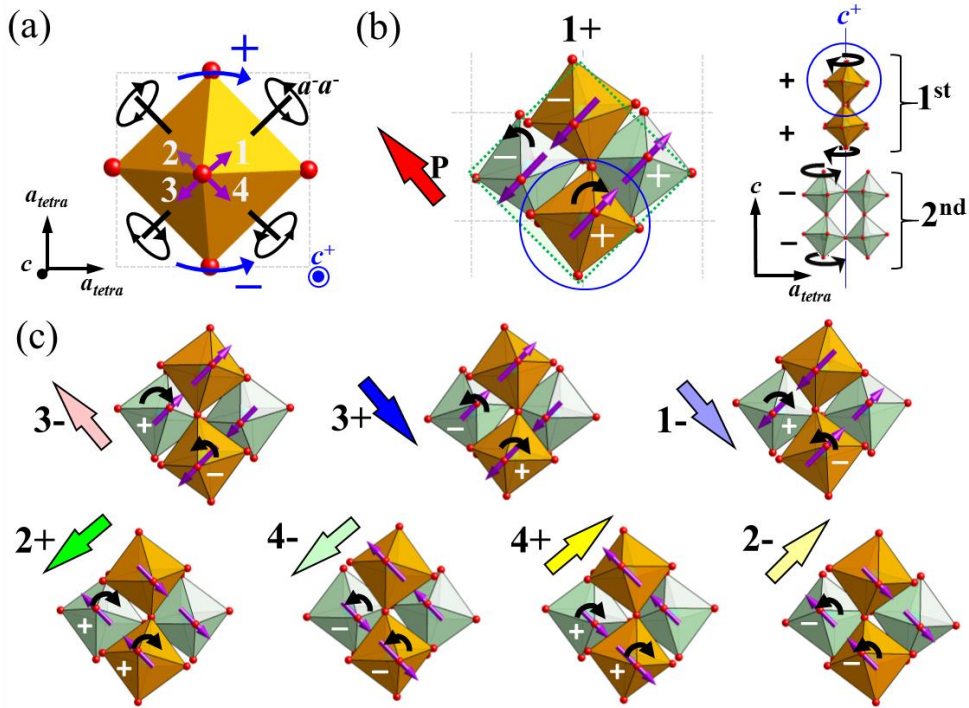

**Figure 1. Structural illustrations of the eight ferroelectric domains in $A_3B_2O_7$. a,** An in-plane $BO_6$ octahedral constituent with tilting along the $<110>_{tetra}$ directions and rotation around the $[001]_{tetra}$ direction. Plus (+) and minus (-) represent clockwise and counterclockwise rotations. Purple arrows indicate the displaced directions ($1^{st}$ to $4^{th}$ quadrants) of the apical oxygen of the $BO_6$ octahedron due to octahedral tilting. **b,** The 1+ domain state. The red arrow indicates the polarization direction. The cyan and red spheres represent B-site and O ions, and A-site cations were omitted for clarity. The gray dashed lines depict the basic tetragonal framework constructed by B-site ions, and the green dotted rectangle depicts the orthorhombic cell. The cross-sectional structure includes two bi-layers formed by orange and light green corner-sharing $BO_6$ octahedra. Each domain state can be unambiguously identified by naming the distortions of a given octahedron (blue circled) since the five adjacent octahedra in each bi-layer are constrained to tilt and to rotate in opposite senses, and also the overall crystallographic symmetry (space group of $A2_1am$) determines the distortions in the adjacent bi-layers. **c,** The in-plane structural models of the 1-, 2±, 3± and 4± domain states. Switching of the octahedral tilting pattern is involved between two domain states in, for example, 1+ vs. 3+, 1- vs. 3-, 2+ vs. 4+, and 2- vs. 4-, and switching of the octahedral rotation pattern is required between 1+ vs. 1-, 2+ vs. 2-, 3+ vs. 3-, and 4- vs. 4+. Antiphase domain relations can be found between two domain states in, for example, 1+ vs. 3-, 1-

vs. 3+, 2+ vs. 4-, and 2- vs. 4+. The corresponding polarization direction in each domain state can be derived readily from our nomenclature; for example, the 1+ domain state accompanies a polarization toward the $2^{nd}$ quadrant— starting from the $1^{st}$ quadrant and rotating clockwise (+) to the $4^{th}$ quadrant that results in the nearest A-site cation (and polarization) being displaced in the opposite direction to the $2^{nd}$ quadrant. Note that the net in-plane dipole moment ($X_5^-$) is caused by this A-site-cation displacement.

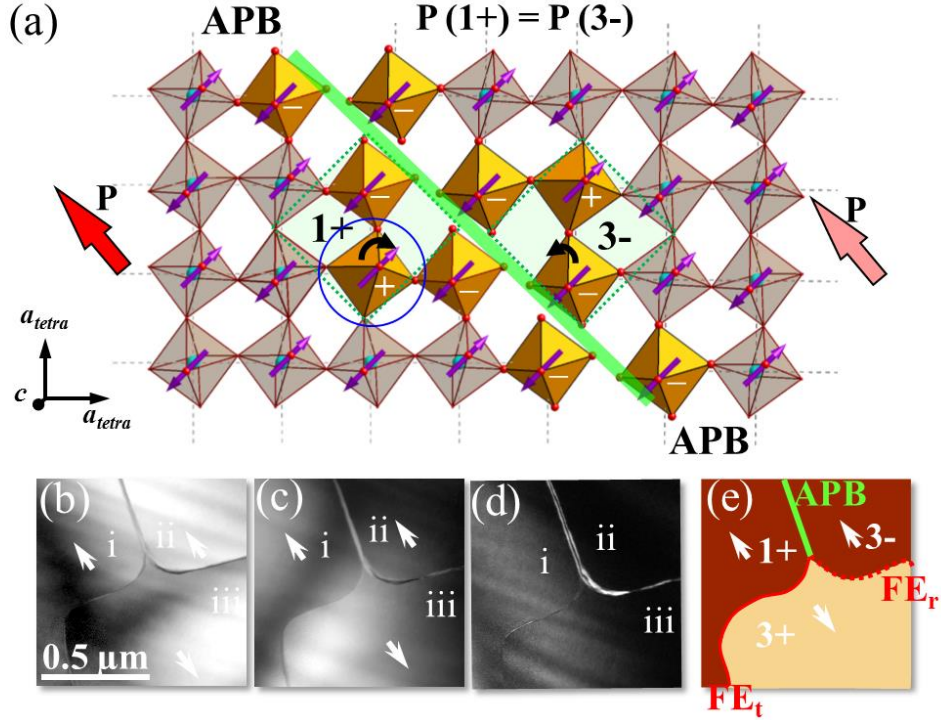

**Figure 2. Antiphase boundary (APB) in HIF A$_3$B$_2$O$_7$. a,** The local distortions near an [110]$_{tetra}$-oriented APB (green line) between the 1+ and 3- states, which are identical in polarization direction but differ in structure with respect to rotation (black curved arrows) and tilting (purple arrows) by 180°. **b,** A *ab*-plane DF-TEM image taken using superlattice g$_1^+$ = 3/2(1, 1, 0)$_{tetra}$ spot. **c,** A DF-TEM image taken using superlattice g$_1^-$ = 3/2(-1, -1, 0)$_{tetra}$ spot. A reversed contrast in (**b**)-(**c**) demonstrates the characteristic of 180°-type FE domains. **d,** A DF-TEM image taken using g$_1^-$ spot at a large tilting angle to tune contrast by enhancing excitation error. A clear boundary interference fringe can then be observed between domains ii & iii, implying an inclined nature and a strong strain gradient expected in rotation-driven FE$_r$ DWs. **e,** The schematic domain configuration obtained from (**b**)-(**d**) demonstrates a typical Z$_3$ vortex pattern within a FA domain, composed of three 180°-FE domains and three DWs: FE$_r$ (red-dotted), FE$_t$ (red-solid) DWs and APB (green-solid). White arrows denote the polarization directions in FE domains.

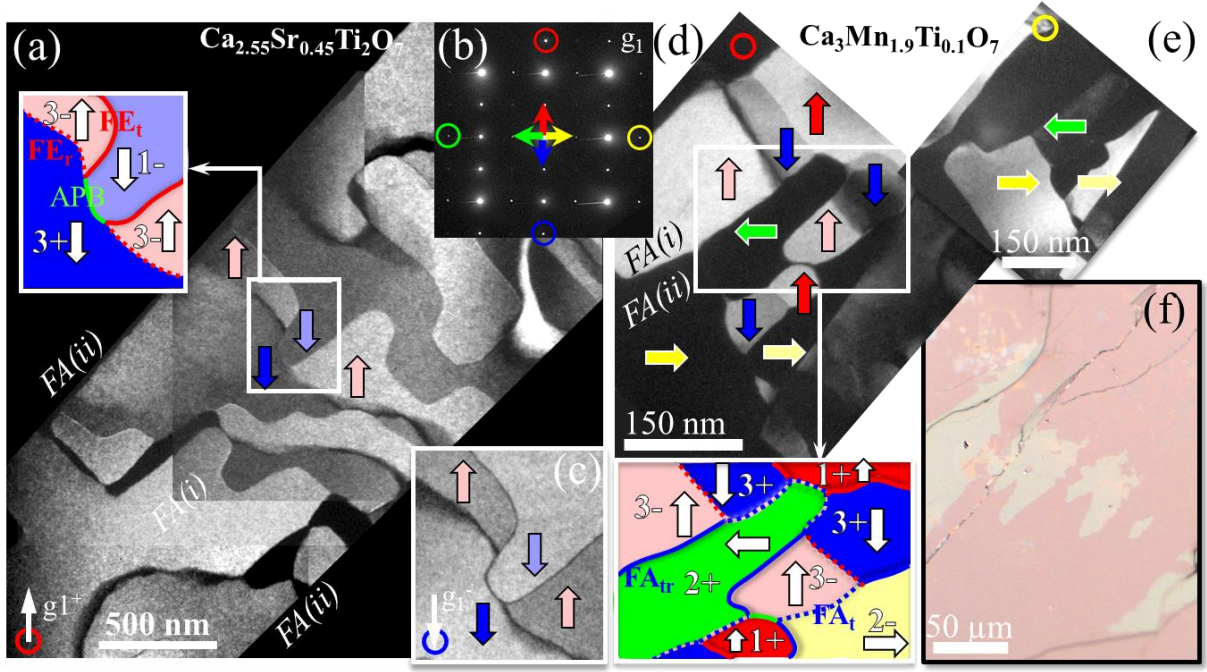

**Figure 3. The Z$_3$-vortex patterns in Ca$_{2.55}$Sr$_{0.45}$Ti$_2$O$_7$ and Ca$_3$Mn$_{1.9}$Ti$_{0.1}$O$_7$ crystals. a,** A 2.5 μm x 2.6 μm mosaic of DF-TEM images were taken using the superlattice g$_1^+$ spot (red-circled) of domain FA(i) in a CSTO crystal along [001]$_{tetra}$. The inset illustrates a proposed domain configuration within the white rectangular box. The colored arrows represent polarization directions within the domain. **b,** The electron diffraction pattern was taken by covering regions FA(i) and FA(ii) showing a 90°-crystallographic-twin relation. The red/blue circled spots were contributed from the orthorhombic distortions of the FA(i) region and the green/yellow ones were from the FA(ii) area. The color-circled spot located in each DF-image is the selected superlattice Bragg spot to light up the corresponding domains at a given orientation. **c,** A DF-TEM image of the white rectangular box region was taken using the blue circled superlattice g$_1^-$ spot. **d-e,** DF-TEM images taken using (**d**) red circle and (**e**) yellow circle spots corresponding to superlattice g$_1^+$ spots in a CMTO crystal. The bottom inset of (**d**) depicts a proposed domain configuration of the white rectangular area in (**d**). A Z$_3$–vortex network appears, with three DWs meeting at one point and with irregular shaped FE/FA domains. Two types of FA DWs, FA$_t$ (blue-dotted) and FA$_{tr}$ (blue-solid), were identified. **f,** Polarized optical image of a CMTO crystal showing irregular twin (i.e. FA) domains in a hundreds-μm scale.

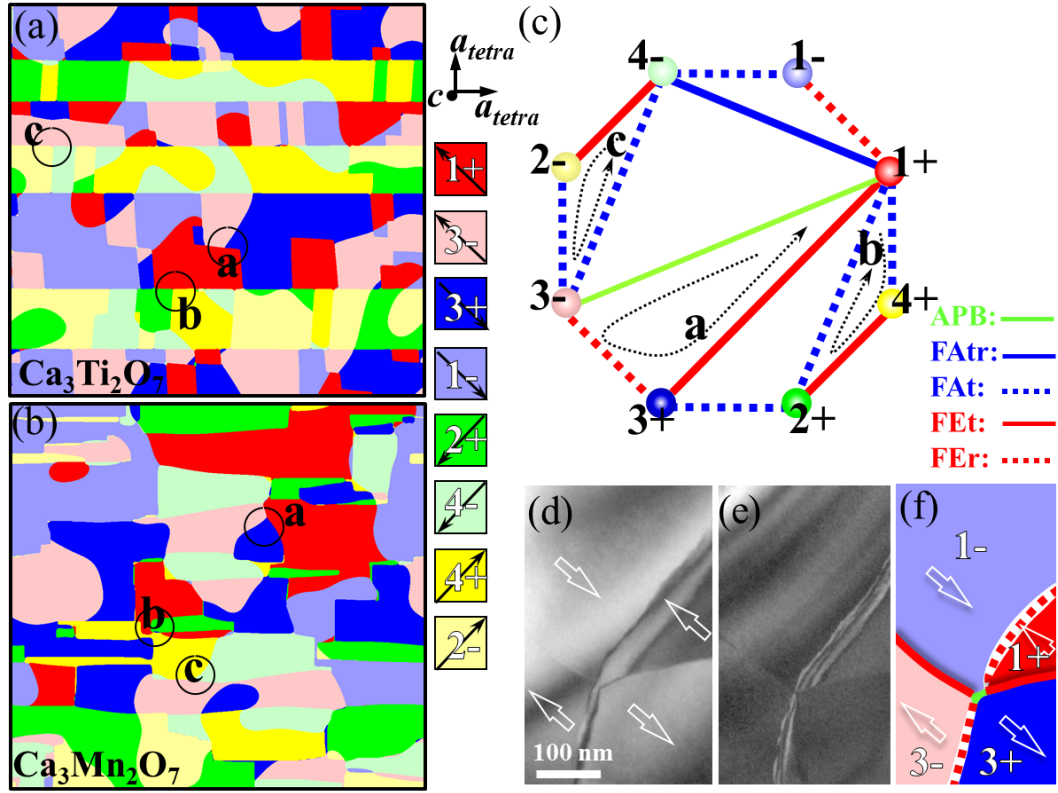

**Figure 4. Z₄xZ₂ domain structures with Z₃ vortices in Ca₃Ti₂O₇ and Ca₃Mn₂O₇. a-b,** In-plane domain structures of Ca₃Ti₂O₇ and Ca₃Mn₂O₇ from phase-field simulations. The eight colors denote the eight domain variants as listed. Z₃ vortices corresponding to loops *a*-c are denoted by black circles in spite of different nature of FA domains in Ca₃Ti₂O₇ and Ca₃Mn₂O₇. **c,** Schematic of the energy diagram in A₃B₂O₇ compounds with 8 vertices, representing eight domain variants. Each vertex is connected to 7 edges that correspond to one of five types of DWs as shown in the right side. Loops a-c depict possible vortex domains and domain walls (in fact, Z₃ vortices). Note that non Z₃-type vortex domains corresponding to loops connecting, for example, (1+, 3+, 3-, 1-, 1+) or (1+, 2+, 3+, 3-, 1+) have not been observed experimentally. **d-f,** Experimental DF-TEM images demonstrate two Z₃ vortices at a very short interval of 50 nm in a Ca₃Ti₂O₇ crystal. Emphasize that the structure is not a Z₄–type vortex. (**d**) was taken under a Friedel's-pair-breaking condition to reveal 180°-type domain contrast. Polarization directions were shown by white arrows. (**e**) was taken under a larger tilting angle to reveal boundary interference fringes clearly. The width of bent fringes indicates a broad DW interrupting the connection of 1+ (red) and 3- (pink) domains. (**f**) shows a schematic of the corresponding domain configuration.

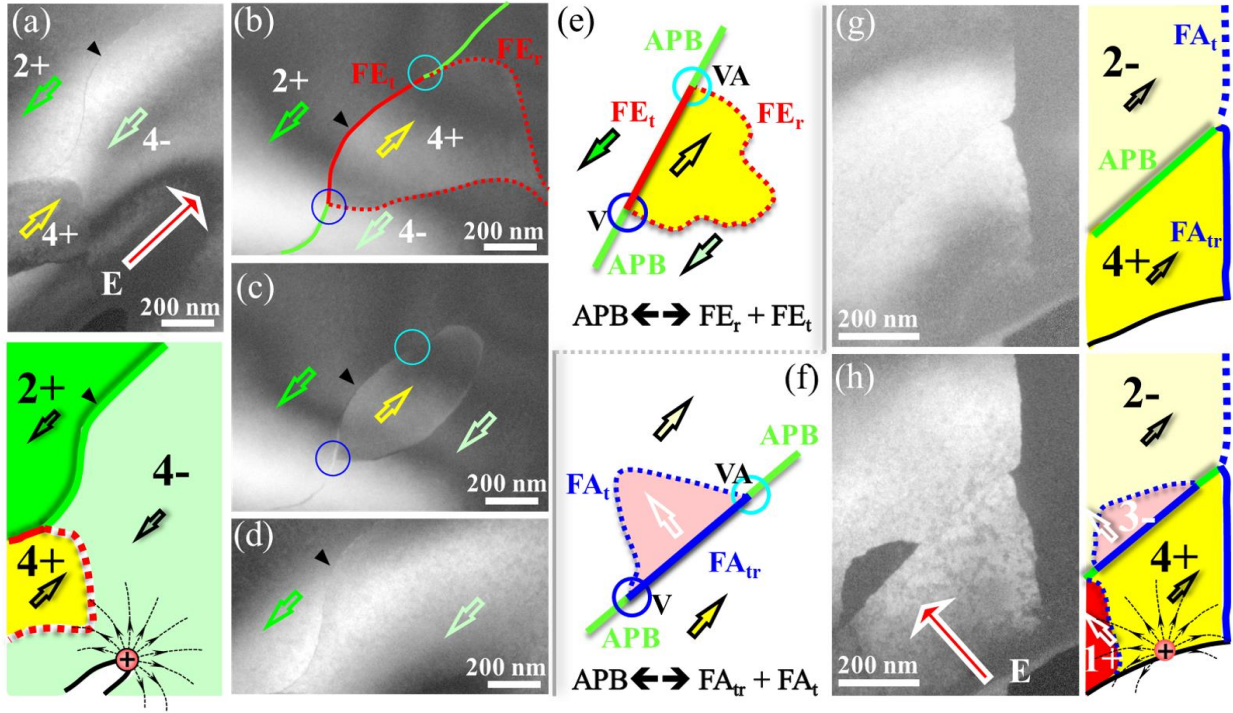

**Figure 5. The domain switching kinetics and vortex-antivortex annihilation/creation under electron beam-induced poling in a CSTO crystal.** In-plane DF-TEM images and schematics of domain structures under different directions of induced electric field indicated by red arrows. Colored arrows and domains represent polarization directions and FE domain states, respectively. **a-d,** Image sequences showing a $Z_3$ vortex-antivortex (V-AV) pair evolution during in-situ poling within a single ferroelastic domain. **a,** The initial state. **b-c,** The states during the charge dissipation process after defocusing electron beam. **d,** The final state. With a focused electron beam at the sample edge, a direct 180º polarization reversal is observed (4-➔4+) via V-AV pair creation ((**a**) to (**b**)), and the created domain disappears slowly with charge dissipation, which accompanies V-AV pair annihilation ((**b**) to (**d**)). Blue and cyan circles denote $Z_3$ vortices and antivortices, respectively. A black arrowhead is the location marker. Emphasize that there is no hint of the presence of any intermediate states corresponding to 90º polarization reversal during this process**. e,** Schematic showing the 180º ferroelectric polarization switching via splitting or coalescence of a APB into two ferroelectric walls: APB (green line) ⬅➔ $FE_t$ (red-solid) + $FE_r$ (red-dotted) DWs. **f,** Schematic showing a 90º ferroelectric polarization switching within a FA domain via splitting or coalescence of a APB into two ferroelastic walls: APB ⬅➔ $FA_{tr}$ (blue-solid) + $FA_t$ (blue-dotted) DWs. **g-h,** Image sequences and schematics showing 90º ferroelectric domain switching

near a APB (green line). **g,** The initial state. **h,** The immediate image after electron beam focused at the sample edge away from the FA boundary (solid blue line). Only $90^o$ poled domains (dark contrast) are observed (2-➔3- and 4+➔1+). The $90^o$ poled domains are assigned with the rotation order parameter same with those of the initial domains, which is consistent with the low-energy nature of $FA_t$ DW. The induced 3- (pink) and 1+ (red) domains return slowly to the initial 2- (light yellow) and 4+ (yellow) states with charge dissipation.

**SI 1: In-plane configuration of ferroelastic and ferroelectric domains in Ca$_{2.55}$Sr$_{0.45}$Ti$_2$O$_7$.**

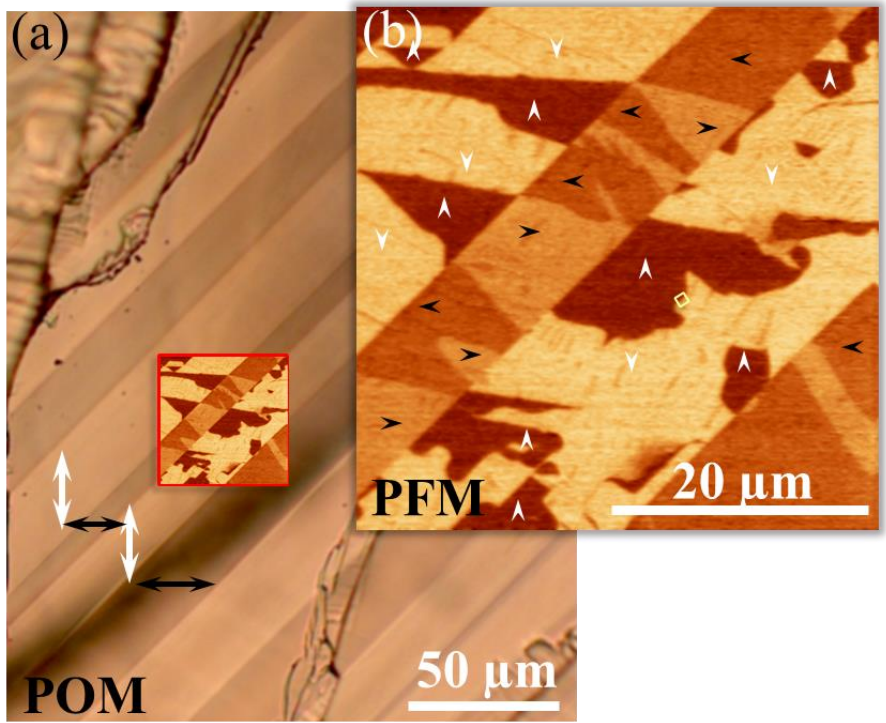

**Supplementary Figure S1. In-plane configuration of ferroelastic and ferroelectric domains in Ca$_{2.55}$Sr$_{0.45}$Ti$_2$O$_7$. a**, A polarized optical microscope (POM) image of a cleaved (001) surface at room temperature. The ferroelastic DWs are [100]$_{tetra}$-oriented. White and black double arrows indicate the polar $a$-axes in an alternating way. **b**, The in-plane piezo-response force microscope (IP-PFM) image of the area in the red box in (**a**) with a scanning area of 40 x 40 μm. The directions of spontaneous in-plane polarization of FE domains are indicated by black and white arrows. There exist four polarization directions along <110>$_{tetra}$ axes.

**SI 2: Local structural distortions at ferroelectric and ferroelastic domain walls.**

Supplementary Fig. S2a shows a (110)$_{tetra}$–oriented FE$_t$ DW between two neighboring 1+ and 3+ domains, where the octahedral tilting (a$^-$a$^-$c$^0$) may be fully suppressed if octahedral tilting changes across the DW by passing through the tetragonal central position (Supplementary Fig. S3). On the other hand, Fig. S2b shows a FE$_r$ DW between neighboring 1+ and 1- domains, where neither of the two lattice modes becomes zero. Thus, FE$_r$ DW may accompany a lower energy than FE$_t$ DW does (Supplementary Fig. S3). Given that the oxygen octahedra in $A_3B_2O_7$ are connected by sharing the oxygen atom in their corner, at those DWs, some shift of equatorial oxygens (indicated by red spheres in black circles in Fig. S2a-b) that locate between Ti sites (cyan spheres) is expected. Enlarged views of the octahedra across the DWs (Fig. S2a-b upper panels) clearly show a larger octahedral mismatch at the $FE_r$ DW than that at the $FE_t$ DW when observed along the $[100]_{tetra}$ axis. Experimentally, APBs can be identified from the domain contrast without ambiguity (Figs. 2b-d in the main text). Although DWs can deviate from typical orientations to minimize the wall energy in the thin foil-type geometry of TEM specimens, we constantly observe a narrow sharp-contrast wall and a relatively broad wall with clear interference fringes near a $Z_3$ vortex core. Since strain provides the main diffraction contrast change in our DF-TEM images, we associate the narrow sharp-contrast lines with a less octahedral mismatch to $FE_t$ DWs in the *ab*-plane projection. This is, indeed, the case for a sharper wall between domain i and iii shown in Fig. 2d in the main text and between domains 1- and 3- or domains 1+ and 3+ shown in Fig. 4e-f in the main text, which we therefore assign as $FE_t$ DWs. A clear interference fringes or wavy features can be observed between domains ii & iii (Fig. 2d in the main text) and domains 1- and 1+ or domains 3- and 3+ (Fig. 4e-f in the main text), suggesting an inclined nature and a strong strain gradient as expected in $FE_r$ DWs.

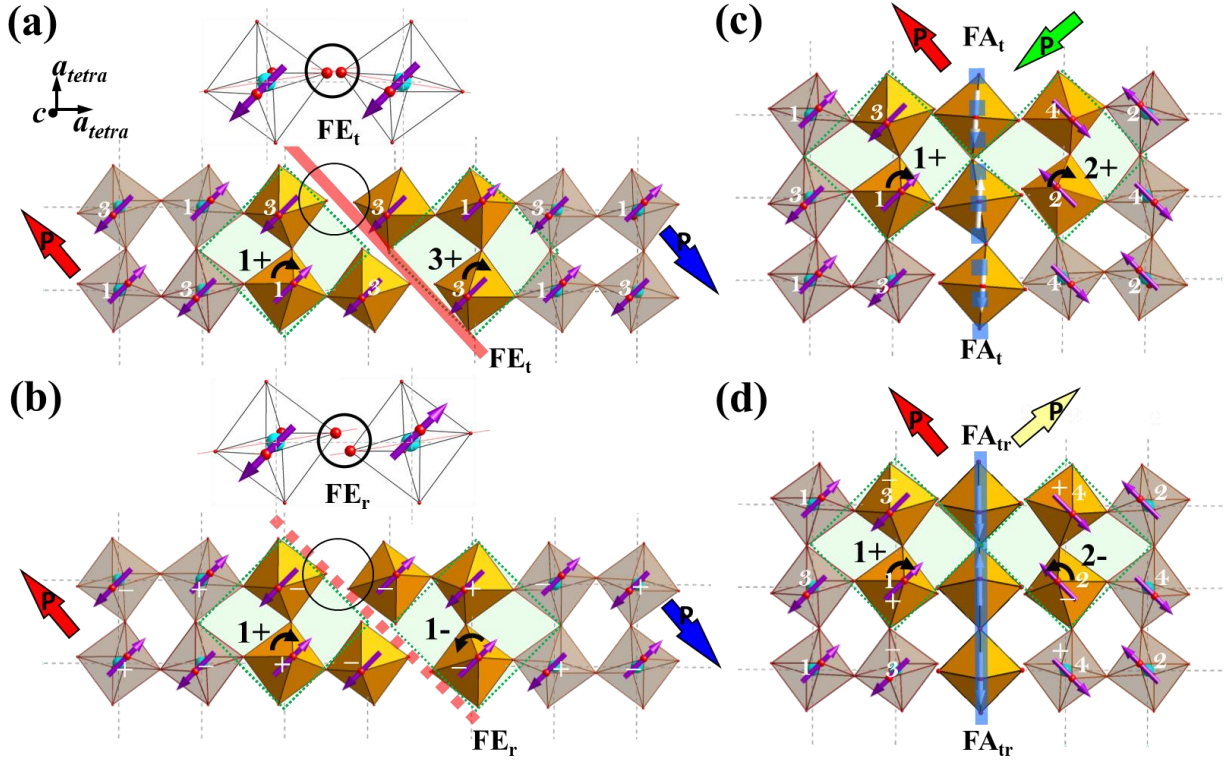

**Supplementary Figure S2. Definition of DWs in A₃B₂O₇.** In-plane illustration of domain walls with BO₆ octahedra represented by orange and gray units. **a,** FE$_t$ DW, red-solid line. A schematic of a (110)$_{tetra}$–oriented FE$_t$ DW between two 180°-type FE domains in 1+ and 3+ states. Green dotted lines for the orthorhombic unit cells demonstrate the discontinuation of octahedral tilting (a⁻a⁻c⁰, purple arrows) across the wall. White Arabic number 1 or 3 marks the displaced direction of the top apical oxygen in each octahedron. **b,** FE$_r$ DW, red-dotted line. A schematic of a (110)$_{tetra}$–oriented FE$_r$ DW between two 180°-type FE domains in 1+ and 1- states. White symbol + (clockwise) or − (counterclockwise) represents the octahedral rotation in each octahedron. **c,** FA$_t$ DW, blue-dotted line. A schematic of a (100)$_{tetra}$–oriented FA$_t$ DW between 90°-type FE domains in 1+ and 2+ states. A residual octahedral distortion of either a⁻a⁰c⁺ or a⁰a⁻c⁺ type at the DW is proposed and shown with white arrows. **d,** FA$_{tr}$ DW, blue-solid line. A schematic of a (100)$_{tetra}$–oriented FA$_{tr}$ DW between 90°-type FE domains in 1+ and 2- states. FA$_{tr}$ DWs may adopt a high-symmetry position with single tilt of either a⁻a⁰c⁰ or a⁰a⁻c⁰ (white arrows) type accompanied by a complete frustration of octahedral rotation at the DW.

**SI 3: Oxygen distortions in eight ferroelectric states.**

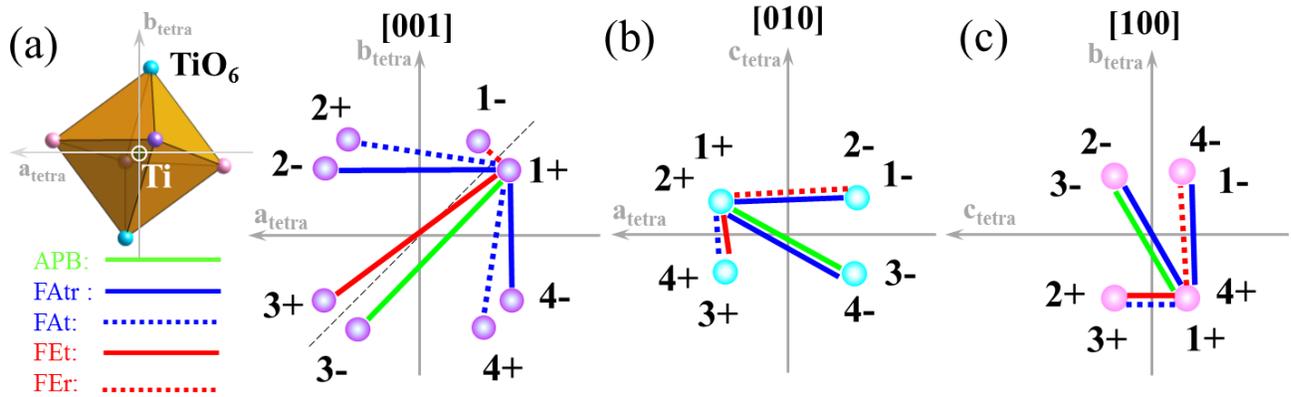

**Supplementary Figure S3. Oxygen positions relative to the tetragonal position in eight ferroelectric states. a,** A view looking down along the *c*-axis of a TiO₆ octahedron in the 1+ state. In order to reveal octahedral distortions in three dimensions, the oxygen positions in 8 states were projected onto **a,** *ab*-plane, **b,** *ac*-plane and **c,** *bc*-plane. By defining Ti-site (white sphere) as the origin (tetragonal position), off-centered oxygen atoms in eight states are depicted. Purple, cyan and pink spheres represent the apical and two equatorial oxygen pairs, respectively. Spheres depict the oxygen distortions including rotation and tilting. The black dashed line between 1+ and 1- (shown in **a**) illustrates the ideal position when only octahedral tilting exists. The deviation of the 1+ and 1- states from the black line results from octahedral rotation. Eight states (spheres) emerge as a result of octahedral tilting and rotation. The solid and dotted lines indicate various DWs linking different states. For example, the green solid line linking 1+ & 3- represents a APB. FA$_t$ DW; the blue dotted line linking 1+ & 2+. FA$_{tr}$ DW; the blue solid line linking 1+ & 2-. FE$_t$ DW; the red solid line linking 1+ & 3+. FE$_r$ DW; the red dotted line linking 1+ & 1-. By assuming that a wall going through a tetragonal oxygen position costs more energy, we can estimate the energy hierarchy among various DWs in three different planar views: (**a**) FE$_r$ < FA$_t$ ~ FA$_{tr}$ < APB < FE$_t$; (**b**) FA$_t$ ~ FE$_t$ < FE$_r$ < FA$_{tr}$ < APB, and (**c**) FA$_t$ < FE$_t$ < FE$_r$ < FA$_{tr}$ < APB. The averaged DW energy hierarchy appears FA$_t$ ≤ FE$_r$ ≤ FE$_t$ < FA$_{tr}$ ≤ APB. Judging from the displaced length of each DWs, APB and FA$_{tr}$ belong to a higher energy set than others. In addition, the oxygen displacement can pass through the origin in APBs, FA$_{tr}$ and FE$_t$ DWs, implying a complete suppression of structural order parameters the middle of the DWs and thus a non-ferroelectric state in the middle of the DWs.

**SI 4: A full assignment of domain states and domain wall types in a Ca$_{2.55}$Sr$_{0.45}$Ti$_2$O$_7$ crystal.**

Supplementary Fig. S4a shows DF-TEM image taken using the superlattice g$_1^-$ spot along the [001] direction, showing a ferroelectric domain contrast opposite that in Fig. 3a in the main text. The domain states and domain wall types are assigned and shown in Fig. S4e, based on our previous discussion on the domain wall features associated with local distortions in the Supplementary information 2. Figures S4b-c show the DF-TEM images taken at the same area as Fig. S4a using the superlattice g$_2^+$ spot along the [1, -1, 1]$_{tetra}$ direction with the specimen 15$^o$ tilted away from the $c$-axis. Non-ferroelectric structural boundaries can be visualized under this condition since the g$_2^+$ vector is perpendicular to the polar $a$-axis and the ferroelectric contribution is minimized. To avoid confusion with APBs, those boundaries observed in this condition without considering the polarization effect are named as "B-boundaries". The appearance of "B-boundaries" is a result of symmetry breaking through the tetragonal-to-orthorhombic phase transition. They tend to show step-like features along the <100>$_{tetra}$ direction (Fig. S4b and Fig. S7f). Based on the domain switching kinetics shown in Fig. 5 in the main text, we argue that those "B-boundaries" are either APBs or FE$_t$ DWs, so APBs and FE$_t$ DWs tend to be <100>$_{tetra}$-oriented. Contrarily, FE$_r$ DWs show wavy features with no preferred orientation (Fig. S4b red-dotted lines), due to the so-called "rotational compatibility conditions" (Ref 38 in the main text). The role of "B-boundaries" is further discussed in Supplementary Fig. S7g.

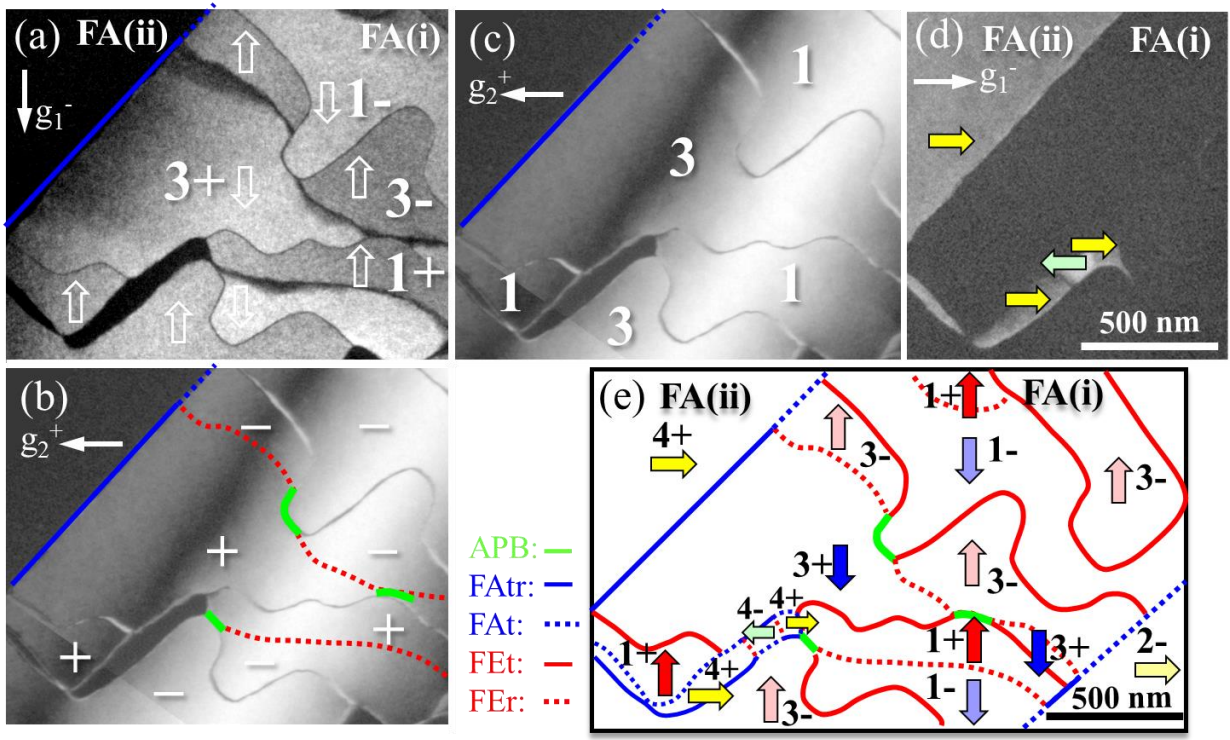

**Supplementary Figure S4. Domain states and domain wall types in a Ca$_{2.55}$Sr$_{0.45}$Ti$_2$O$_7$ crystal.**
**a,** A DF-TEM image taken using the superlattice g$_1^-$ = 3/2(-1, -1, 0)$_{tetra}$ = (-3, 0, 0)$_{orth}$ along the
[001] direction, showing a domain contrast opposite to that in Fig. 3a in the main text. **b,** A DF-
TEM image taken using the superlattice g$_2^+$ = 3/2(-1, 1, 2)$_{tetra}$ = (0, -3, 3)$_{orth}$ along [1, -1, 1]$_{tetra}$ with
specimen 15$^o$ tilted away from the *c*-axis. Only a part of DWs in (a) are observed in (b). Invisible
red-dotted FE$_r$ DWs are depicted, based on the broad wavy features shown in (a). Green APBs are
also shown based on the domain contrast shown in (a). It also shows a possible rotation
configuration where opposite rotations (+ and −) are separated by red-dotted FE$_r$ DWs. **c,** A tilting
(1$^{st}$ or 3$^{rd}$ quadrants) configuration of (b). **d,** A DF-TEM image taken using another superlattice
g$_1^+$ spot, 90$^o$ relative to the one for FA(i). The bright contrast corresponds to the other four types
of 180$^o$ FE domains (2+, 2-, 4+ and 4-) inside the FA(ii) region. **e,** A schematic domain and domain
wall configuration for the area in Fig. 3a in the main text and Fig. S4a, including various FE
domains and five types of DWs.

**SI 5: In-plane configuration of ferroelastic and ferroelectric domains in Ca₃Mn₁.₉Ti₀.₁O₇.**

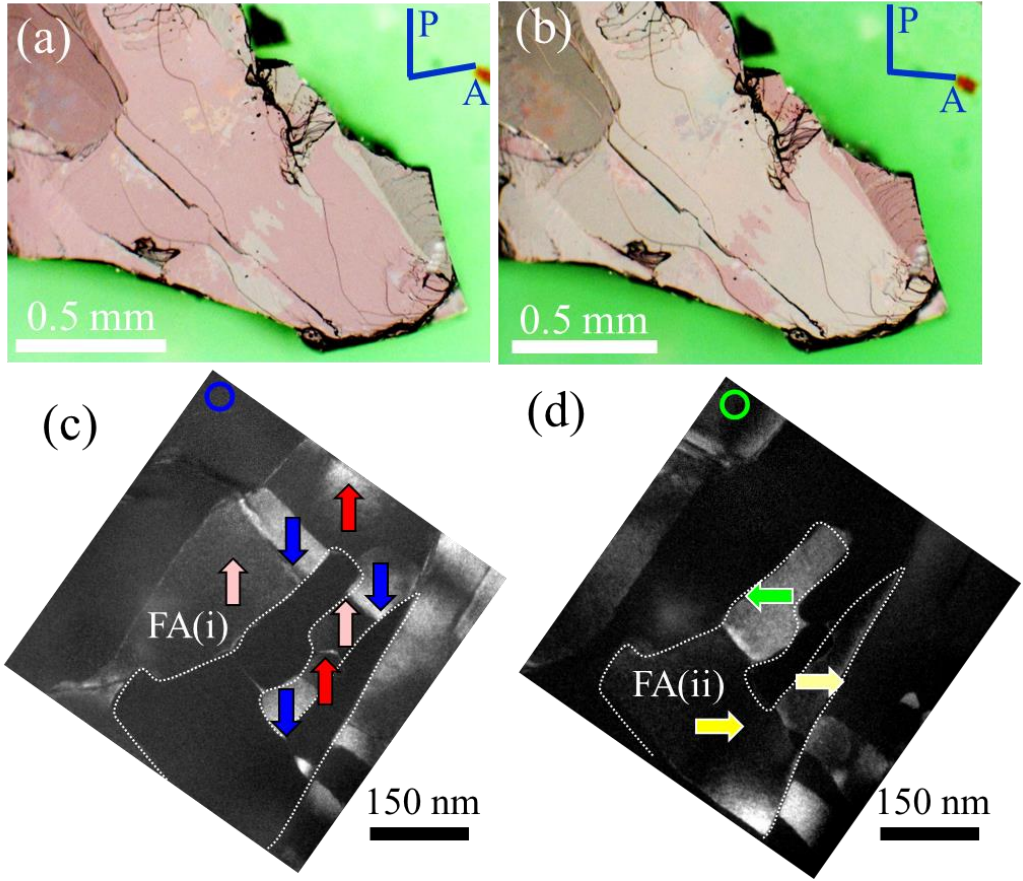

**Supplementary Figure S5. In-plane configuration of ferroelastic and ferroelectric domains in a Ca₃Mn₁.₉Ti₀.₁O₇ crystal. a-b,** The linearly polarized optical microscope (POM) images on the cleaved (001) surface of a Ca₃Mn₁.₉Ti₀.₁O₇ crystal. The angle between polarizer (P) and analyzer (A) is shown in the figure. Contrast reversal of POM images with the change of analyzer angle and an orthogonal relation of superlattice spots in electron diffraction patterns both confirm those irregular domains as the 90°-oriented FA domains (i.e. orthorhombic twins). **c-d,** DF-images were taken using the $g_1^-$ spot, circled in blue or green in Fig. 3b in the main text. Colored arrows correspond to some of eight FE states, and white dotted lines depict FA DWs between FA(i) and FA(ii).

**SI 6: Five types of Z₃-vortex configurations.**

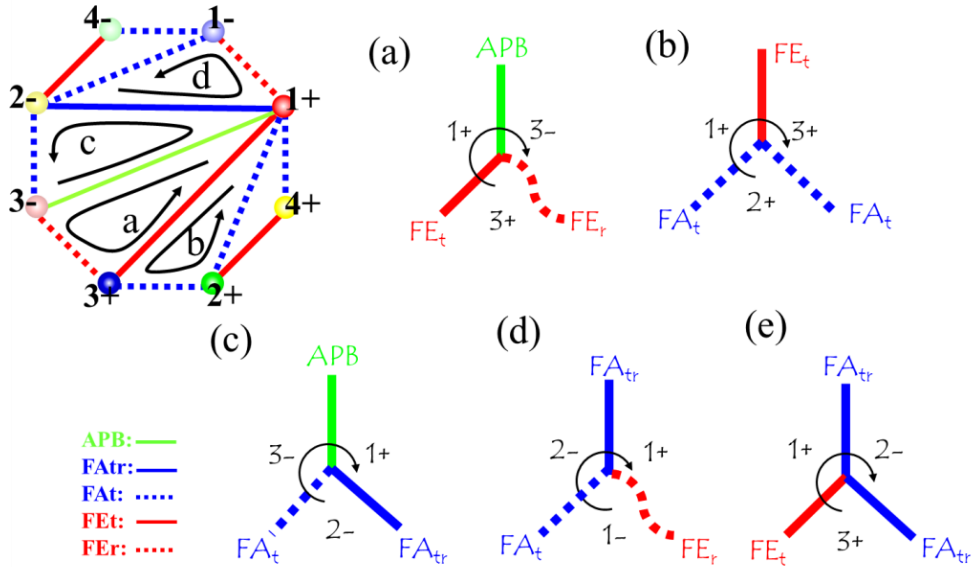

**Supplementary Figure S6. Z₃-vortex configurations in HIF A₃B₂O₇.** APB; green solid line, FA$_{tr}$; blue-solid line, FA$_t$; blue-dotted line, FE$_t$ DW; red-solid line, and FE$_r$; red-dotted line. The indicated domain states are examples. **a,** The only Z₃-vortex configuration appearing within a single orthorhombic twin. This type Z₃ vortex has been frequently observed and discussed extensively in the main manuscript (see, for example, Fig. 2b). **b,** The most common Z₃ vortex across the orthorhombic twin boundaries. This type Z₃ vortex is, for example, shown in Fig. 3d in the main manuscript. **c,** Z₃ vortex accompanying 90º ferroelectric switching at APB, as shown in Fig 5h in the main manuscript. **d,** Z₃ vortex accompanying ferroelectric switching in the absence of APB, as shown in the Supplementary Fig. S8. **e,** The least favored Z₃ vortex with two high energy FA$_{tr}$ walls. This type Z₃ vortex has also been experimentally observed, as shown in the Supplementary Fig. S7j.

**SI 7: Polarization domain switching kinetics under electron beam-induced poling in the presence of APBs.**

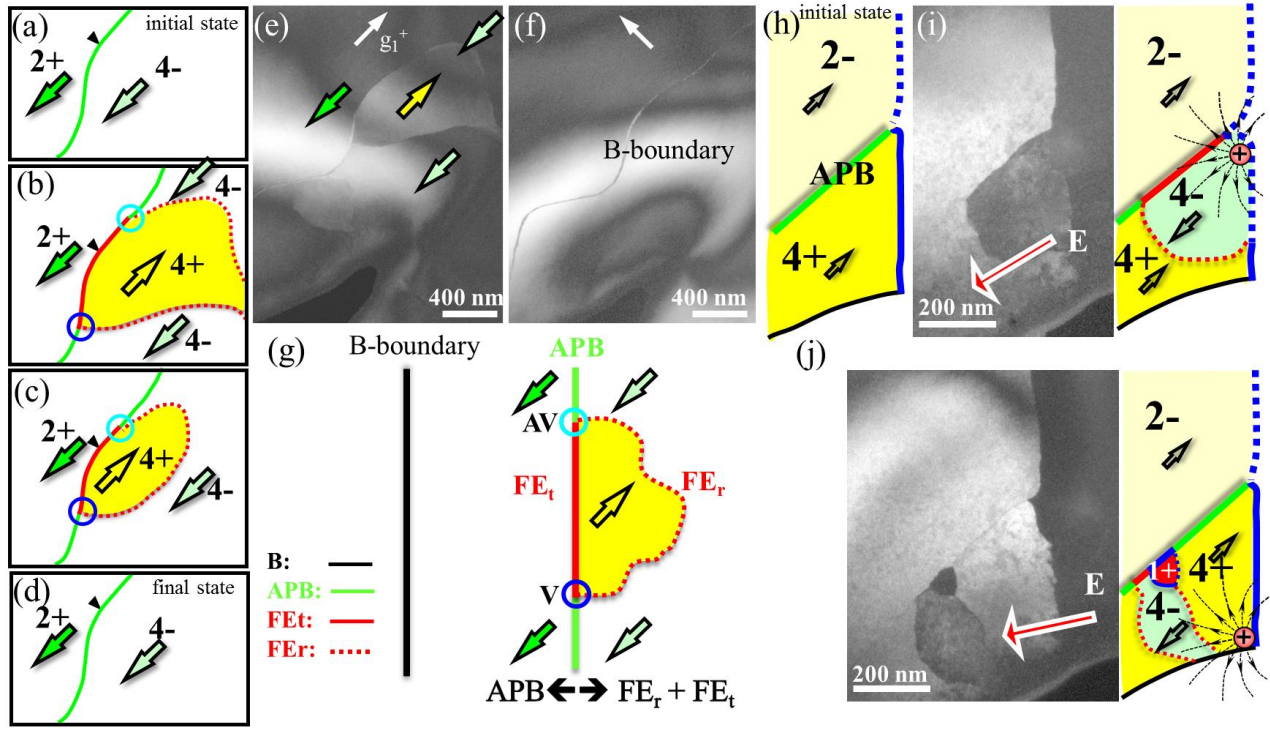

**Supplementary Figure S7. Polarization domain switching kinetics close to APBs in a Ca$_{2.55}$Sr$_{0.45}$Ti$_2$O$_7$ crystal, showing Z$_3$ vortex-antivortex creation and annihilation. a-d,** In-plane schematics of beam-induced poling effects correspond to Fig. 5a-d in the main text. Color arrows represent different FE domain states. Blue and cyan circles denote Z$_3$ vortex (V) and antivortex (AV), respectively. APBs: green solid line; FE$_t$ DWs: red solid line, and FE$_r$ DWs: red dotted line. **e,** A superlattice DF-TEM image taken using the g$_1^-$ spot, corresponding to (**b**). **f,** A superlattice DF-image acquired on the same area as (**e**) using the g$_2^+$ spot shows a "B-boundary". In this condition, no FE domain contrast is revealed. **g,** A schematic of a DW evolution during a 180º ferroelectric polarization switching via a splitting or coalescence between APB(tr) $\longleftrightarrow$ FE$_t$ + FE$_r$ DWs. Note that APBs are associated with simultaneous discontinuities of octahedral rotation (r) and tilting (t) at the boundaries, compared with tilting (rotation) discontinuity at FE$_t$ (FE$_r$) DWs. APB (tr) can split into one FE$_t$ DW and one FE$_r$ DW, and this split accompanies a Z$_3$ V-AV pair creation as seen from Fig. S7a to b. On the other hand, a V-AV pair annihilation occurs with the coalescence of one FE$_t$ DW and one FE$_r$ DW into one APB as seen from Fig. S7b to d. Our results demonstrate that FE$_r$ DWs can move readily with an applied electric field while FE$_t$ DWs and APBs are attached to "B-boundaries". B-boundaries remain intact during ferroelectric switching, implying that all B-boundaries become APBs in the fully poled state. **h-j,** Electron beam-induced evolutions of FE domains. **h,** The initial state. **i,** With electron beam focused at the center of a FA boundary, a 180° polarization reversal (dark gray contrast) is observed (4+➔4-). The beam-induced 4- domain returns slowly back to the initial 4+ state with charge dissipation. **j,** Consecutively, electron beam was focused at the bottom of the FA boundary. A major 180° polarization reversal (4+➔4-) and a minor 90° switching (4+➔1+) are observed. Again, the beam-induced domains return slowly back to the initial state. Only 90° poled domains (dark contrast) are observed when the electron beam was consecutively focused at the sample edge away from the FA boundary (Fig. 5h in the main text). Our results demonstrate that the switching process depends significantly on the electric field orientations.

**SI 8: Polarization domain switching kinetics under electron beam-induced poling in the absence of APBs.**

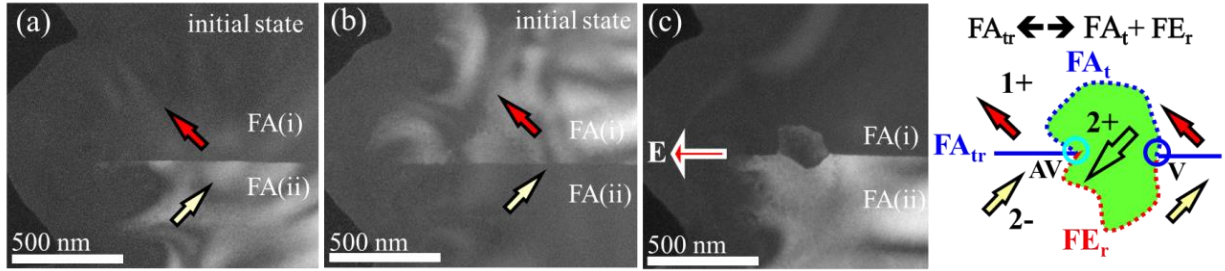

**Supplementary Figure S8. Polarization domain switching kinetics close to a FA wall and in the absence of APBs in a Ca$_{2.55}$Sr$_{0.45}$Ti$_2$O$_7$ crystal. a-b,** In-plane DF-TEM images show two ferroelastic (FA(i) and FA(ii)) regions. Domain states 2- (light yellow arrow) and 1+ (red arrow) are assigned based on diffraction patterns and the assumption of a non-charged head-to-tail wall. **c,** Polarization domain evolutions across a FA$_{tr}$ boundary. Election beam focused at the end of a FA boundary close to the sample edge induces an electric field indicated by a red/white arrow. A polarization switching (dark gray contrast) from 1+ ➔ 2+ in FA(i) and 2- ➔ 2+ in FA(ii) is observed. The right-side cartoon shows a schematic of the domain switching via a splitting/merging of FA$_{tr}$ ⬌ FA$_t$ + FE$_r$ DWs. A high energy FA$_{tr}$ DW becomes two DWs with lower energies; one ferroelectric FE$_r$ DW and the other ferroelastic FA$_t$ DW, which is consistent

with our energy hierarchy. Both $FA_t$ and $FE_r$ DWs tend to be mobile and highly curved, which is again consistent the low energy nature of $FA_t$ and $FE_r$ DWs in our energy hierarchy. In comparison, newly-formed $FE_t$ and $FA_{tr}$ DWs from APBs are pinned preferentially at the original APB locations as shown in Fig. 5e-f in the main text. Blue and cyan circles denote $Z_3$ vortex (V) and antivortex (AV), respectively.

## SI 9: Coefficients of $Ca_3Ti_2O_7$ used in the phase-field simulations

Total energy calculations based on density functional theory (DFT) within the generalized-gradient approximation given by the revised Perdew-Becke-Erzenhof (PBEsol) parameterization for solids [Ref 43 in the main text] using the projector augmented wave (PAW) method [Refs 44-45 in the main text] implemented in the Vienna *Ab Initio* Simulation Package (VASP) [Refs 46-47 in the main text] are used to obtain the coefficients found in the Landau polynomial (Eq. 1, Methods). A plane-wave cutoff of 600 eV and a 4×4×1 *k*-point mesh with Gaussian smearing (0.10 eV width) is used for the Brillouin-zone integrations. The calcium $3s$, $3p$, and $4s$ electrons, Ti $3p$, $3d$, and $4s$ electrons, and O $2s$ and $2p$ electrons are treated as valence states.

The coefficients for the Landau polynomial are obtained by fitting the calculated total energies as a function of magnitude of the order parameters for the configurations corresponding to displacement patterns for each individual mode or combination of modes. The values for the relevant coefficients are given in Supplementary Table 1. In all total energy calculations, the lattice constants are fixed at the calculated equilibrium values for the high symmetry $I4/mmm$ structure (parent clamping approximation). The total elastic stiffness tensor, including the contributions for distortions with rigid ions and the contributions from relaxed ions, is also obtained by calculating the strain-stress relations [Ref 48 in the main text] in the $I4/mmm$ structure (see Supplementary Table 2). The gradient energy coefficients $\kappa_{ijkl}$, $\delta_{ijkl}$, and $g_{ijkl}$ are estimated based on the gradient energy coefficients of $BiFeO_3$ [Ref 38 in the main text] since both the two systems show the coexistence of oxygen octahedral tilt and polarization, and are listed in Supplementary Table 3. Note that the domain structures are determined by the relative magnitude of different gradient energy coefficients, and will be hardly affected by the specific values.

**Supplementary Table 1. DFT calculated coefficients for the Landau polynomial.**

| Coefficient | $\alpha_{33}$ | $\alpha_{3333}$ | $\beta_{11}$ | $\beta_{1111}$ | $\beta_{1122}$ | $\gamma_{11}$ | $t_{3311}$ | $d$ |
|---|---|---|---|---|---|---|---|---|
| Unit | eV f.u.$^{-1}$ Å$^{-2}$ | eV f.u.$^{-1}$ Å$^{-4}$ | eV f.u.$^{-1}$ Å$^{-2}$ | eV f.u.$^{-1}$ Å$^{-4}$ | eV f.u.$^{-1}$ Å$^{-4}$ | eV f.u.$^{-1}$ Å$^{-2}$ | eV f.u.$^{-1}$ Å$^{-4}$ | eV f.u.$^{-1}$ Å$^{-3}$ |
| Value | -0.3505 | 0.1868 | -0.2024 | 0.0527 | 0.0176 | 0.0188 | 0.1464 | -0.1701 |

**Supplementary Table 2. DFT calculated elastic stiffness tensor coefficients.** All values are given in units of GPa.

| Coefficient | $C_{11}$ | $C_{12}$ | $C_{13}$ | $C_{33}$ | $C_{44}$ | $C_{66}$ |
|---|---|---|---|---|---|---|
| Value | 315 | 77.6 | 98.0 | 297 | 81.6 | 83.1 |

**Supplementary Table 3. Normalized gradient energy coefficients.** All values are normalized with respect to $g_{110}$=2.2 eV f.u.$^{-1}$.

| Coefficient | $\kappa_{1111}$ | $\kappa_{1122}$ | $\kappa_{1122}$ | $\delta_{1111}$ | $\delta_{1122}$ | $\delta_{1212}$ | $g_{1111}$ | $g_{1122}$ | $g_{1212}$ |
|---|---|---|---|---|---|---|---|---|---|
| Value | 0.88 | -8.8 | 8.8 | 0.88 | -8.8 | 8.8 | 0.50 | -0.088 | 0.088 |